# An infrared divacancy color center in bulk hexagonal diamond for quantum communication

Xuanxuan Zhang,[1,2] Hanning Pei,[1,2] Shuxin Xu,[1,2] Guliqinayi Alimu,[1,2] Cong Li,[1,2,*] Chen-Min Dai,[1,2,*] Chunlan Ma,[1,2] and Zenghua Cai[1,2,*]

[1]*Key Laboratory of Intelligent Optoelectronic Devices and Chips of Jiangsu Higher Education Institutions, School of Physical Science and Technology, Suzhou University of Science and Technology, Suzhou, 215009, China*

[2]*Advanced Technology Research Institute of Taihu Photon Center, School of Physical Science and Technology, Suzhou University of Science and Technology, Suzhou, 215009, China*

[*]Corresponding authors: cong.li@usts.edu.cn; daichenmin@usts.edu.cn; zhcai@usts.edu.cn

**Abstract**

We investigate intrinsic color centers in bulk hexagonal diamond as potential emitters for fiber-based quantum communication. Hybrid density functional theory and constrained-occupation ΔSCF calculations identify five carbon-vacancy and divacancy configurations with stable optical excited states. Comparing their radiative and nonradiative properties within a three-point configuration-coordinate model identifies the positively charged axial divacancy, $VV_a^{1+}$, as the most promising candidate. It has the highest calculated zero-phonon-line (ZPL) emission rate (43.9 MHz) and ZPL quantum efficiency (11.8%) among the five candidates. Its weak structural relaxation gives a Huang-Rhys factor of 0.395, while a multimode photoluminescence calculation predicts a Debye-Waller factor of 69.37% and a dominant zero-phonon contribution. Its predicted ZPL at 0.603 eV (2057.3 nm) is also closest in photon energy to the 1310 and 1550 nm telecom targets. Quantum frequency upconversion could connect this strong infrared ZPL to the telecom O and C bands, providing a route toward fiber-based quantum communication with an intrinsic color center.

## I. INTRODUCTION

Color centers in cubic diamond provide solid-state platforms for quantum sensing and quantum networks. The negatively charged nitrogen-vacancy center, $NV^{1-}$, combines optical spin initialization and readout with long spin coherence, enabling nanoscale magnetometry and entanglement distribution between remote nodes.[1-3] The negatively charged silicon-vacancy center, $SiV^{1-}$, offers a complementary optical interface through strong zero-phonon-line (ZPL) emission and highly indistinguishable photons. SiV-based quantum memories have also been entangled through deployed telecom fiber.[4, 5] These advances motivate the search for color centers with favorable spin properties and efficient emission into a well-defined optical transition. The ZPL energy, transition dipole moment, and electron–phonon coupling determine the emission wavelength, radiative rate, and competition between ZPL emission, phonon-sideband emission, and nonradiative decay.[6]

Hexagonal diamond, also known as lonsdaleite, is an alternative carbon host for color centers. Both diamond polymorphs consist of tetrahedrally bonded carbon, but their different stacking sequences produce distinct crystal fields and defect symmetries. Recent synthesis of bulk hexagonal diamond and subsequent preparation of millimeter-scale phase-pure material have made this polymorph increasingly accessible to experiment.[7, 8] These developments complement predictions of a higher indentation strength than cubic diamond and measurements of exceptional hardness and thermal stability.[7, 9] The structural differences modify defect orbitals and their coupling to lattice vibrations, potentially producing optical properties distinct from those of analogous defects in cubic diamond.

The optical properties of defects in hexagonal diamond are only beginning to be explored. First-principles studies have investigated the electronic structure and luminescence of NV centers in bulk hexagonal diamond, and recent work has examined their spin coherence.[10-12] Our previous study characterized intrinsic defects, impurities, and defect complexes in hexagonal diamond and identified vacancy-related defects as potential color centers.[13] Assessing their suitability as optical emitters requires analysis of the relaxed excited states and the radiative and nonradiative transitions following optical excitation. A defect-level diagram alone does not determine the ZPL emission strength or the fraction of excited-state population that decays radiatively.

Here, we investigate carbon vacancies and two nearest-neighbor divacancy configurations using hybrid density functional theory and constrained-occupation ΔSCF calculations. We evaluate their optical properties with the three-point configuration-coordinate framework of Iwamoto *et al.*, which combines excited-state relaxation with radiative and multiphonon nonradiative transitions.[14] Comparing ZPL emission rates, quantum efficiencies, and proximity to telecom wavelengths identifies the positively charged axial divacancy, $VV_a^{1+}$, as the most promising intrinsic infrared emitter. We calculate its photoluminescence (PL) spectrum to resolve the phonon sideband and determine its Debye–Waller factor (DWF). Its emission lies outside the conventional telecom windows, so we consider quantum frequency conversion (QFC) as a route to interfacing it with telecom photonics.

## II. COMPUTATIONAL METHODS

### A. Electronic structure and defect models

We performed spin-polarized density functional theory calculations using VASP and the projector augmented-wave method.[15, 16] All defect configurations were constructed in a 6 × 6 × 4 supercell (576 atoms) of the four-atom primitive cell of 2H diamond. The plane-wave energy cutoff was 420 eV, and the Brillouin zone was sampled at the Γ point. The electronic energy convergence threshold was $10^{-5}$ eV, and atomic positions were relaxed until the forces were below 0.01 eV/Å.

Both structural relaxation and total-energy calculations employed the screened Heyd–Scuseria–Ernzerhof (HSE) hybrid functional.[17] An exact-exchange fraction of 0.28 gives a band gap of 4.45 eV, in good agreement with the GW value of 4.68 eV.[18] The ground- and excited-state geometries used in the optical calculations were obtained with this functional.

We consider intrinsic point defects and vacancy complexes in hexagonal diamond: the carbon vacancy, $V_C$, and nearest-neighbor divacancies in the axial and basal configurations, denoted $VV_a$ and $VV_b$, respectively. The charge states include all thermodynamically stable states identified in our previous calculations as the Fermi level varies across the band gap: $V_C^0$, $V_C^{1+}$, and $V_C^{1-}$ for the carbon vacancy, and the neutral, singly positive, singly negative, and doubly negative charge states for both divacancy configurations.[13]

### B. Excited states and the three-point configuration-coordinate model

Optical excited states were calculated using the constrained-occupation ΔSCF method.[19] An electron was promoted from an occupied to an unoccupied defect state within the same spin channel, and the atomic positions were relaxed under the imposed occupation. The ordering and occupation of the defect levels were examined after relaxation to identify interchanges between the occupied and unoccupied states involved in the transition.[20, 21]

The optical properties were evaluated using the three-point configuration-coordinate model.[14] In Fig. 1, $Q_g$ and $Q_e$ denote the equilibrium configuration coordinates of the ground and excited states. Points A, B, and C correspond to the relaxed ground state, the relaxed excited state, and the ground-state electronic configuration at the excited-state geometry, respectively. Their total energies define the ZPL energy and the ground-state Franck–Condon relaxation energy,

$$E_{\mathrm{ZPL}} = E_B - E_A, \quad E_{\mathrm{FC}} = E_C - E_A, \qquad (1)$$

where $E_A = E_g(Q_g)$, $E_B = E_e(Q_e)$, and $E_C = E_g(Q_e)$. The vertical absorption and emission energies are $E_{\mathrm{abs}} = E_e(Q_g) - E_g(Q_g)$ and $E_{\mathrm{em}} = E_B - E_C$, respectively. The mass-weighted displacement between the two equilibrium structures is

$$\Delta Q = Q_e - Q_g = \sqrt{\sum_{\alpha,i} m_\alpha (r_{e,\alpha i} - r_{g,\alpha i})^2}, \qquad (2)$$

where $m_\alpha$ is the mass of atom α and $r_{g,\alpha i}$ and $r_{e,\alpha i}$ are its Cartesian coordinates in the ground and excited states; i runs over x, y, and z. The ground- and excited-state potential-energy surfaces are approximated by parabolas with the same effective frequency, $\Omega_g = \Omega_e = \Omega$,

$$E_g(Q) = E_A + \tfrac{1}{2}\Omega^2 (Q - Q_g)^2, \quad E_e(Q) = E_B + \tfrac{1}{2}\Omega^2 (Q - Q_e)^2. \qquad (3)$$

The three total energies and ΔQ then determine Ω, the Huang–Rhys factor S, and the DWF:

$$\Omega = \sqrt{\frac{2E_{\mathrm{FC}}}{(\Delta Q)^2}}, \quad S = \frac{\Omega(\Delta Q)^2}{2\hbar} = \frac{E_{\mathrm{FC}}}{\hbar\Omega}, \quad DWF = e^{-S}. \qquad (4)$$

The DWF is the squared overlap of the ground vibrational states of the two electronic configurations (W in Ref. [14]). A small S indicates weak electron–phonon coupling and a large ZPL contribution.

The electric-dipole radiative rate before including the phonon-induced reduction in emission energy is

$$\Gamma_{\mathrm{R}}^0 = \frac{n E_{\mathrm{ZPL}}^3 |\mu|^2}{3\pi\varepsilon_0 \hbar^4 c^3}, \qquad (5)$$

where μ is the transition dipole moment (see Ref. [14] for the definition) between the participating defect orbitals, n = 2.4 is the refractive index, $\varepsilon_0$ is the vacuum permittivity, ℏ is the reduced Planck constant, and c is the speed of light. The local-field correction is set to unity, as in Refs. [6] and [14]. At low temperature, the total radiative rate $\Gamma_R$, ZPL rate $\Gamma_{\mathrm{ZPL}}$, and phonon-sideband rate $\Gamma_{\mathrm{PSB}}$ are evaluated as

$$\Gamma_R = \Gamma_R^0 \left(1 - \frac{S\hbar\Omega}{E_{ZPL}}\right), \quad \Gamma_{ZPL} = \Gamma_R^0 \, DWF, \quad \Gamma_{PSB} = \Gamma_R - \Gamma_{ZPL}. \tag{6}$$

Nonradiative relaxation is treated as a multiphonon transition from the lowest vibrational level of the excited electronic state. In the static-coupling approximation, its rate is

$$\Gamma_{NR} = \frac{2\pi}{\hbar} g w_{e,g}^2 \sum_j |\langle \chi_{e,0}|\hat{Q}|\chi_{g,j}\rangle|^2 \, \delta(E_{ZPL} - j\hbar\Omega), \tag{7}$$

where g is the configurational degeneracy factor, $\hat{Q}$ is the configuration-coordinate operator, $w_{e,g}$ is the electron–phonon coupling matrix element, and $\chi_{e,0}$ and $\chi_{g,j}$ are the excited-state vibrational ground state and the jth ground-state vibrational level. Following the analytical treatment of Ref. [14], we introduce $\nu = E_{ZPL}/(S\hbar\Omega)$ and obtain the continuous-level approximation

$$\Gamma_{NR} = \frac{2\pi}{\hbar} g w_{e,g}^2 \frac{(\Delta Q)^2}{\hbar\Omega} \frac{e^{-S} S^{\nu S}}{\Gamma(\nu S + 1)} \times \left(\frac{\nu}{2} + \sqrt{2}\right)\left(\frac{\nu}{2} + \sqrt{2} - 1\right). \tag{8}$$

Here Γ(ν + 1) is the Euler gamma function. We use g = 3 for $VV_a^{2-}$ and g = 1 for the other four centers, with the representative coupling $w_{e,g}$ = 0.1 eV amu$^{-1/2}$ Å$^{-1}$ adopted in Ref. [14]. The quantities $E_{ZPL}$, ΔQ, Ω, and μ are evaluated separately for each defect. The radiative and ZPL quantum efficiencies are

$$\eta_{rad} = \frac{\Gamma_R}{\Gamma_R + \Gamma_{NR}}, \quad \eta_{ZPL} = \frac{\Gamma_{ZPL}}{\Gamma_R + \Gamma_{NR}}, \tag{9}$$

and the corresponding radiative and total excited-state lifetimes are

$$\tau_{rad} = \frac{1}{\Gamma_R}, \quad \tau_{total} = \frac{1}{\Gamma_R + \Gamma_{NR}}. \tag{10}$$

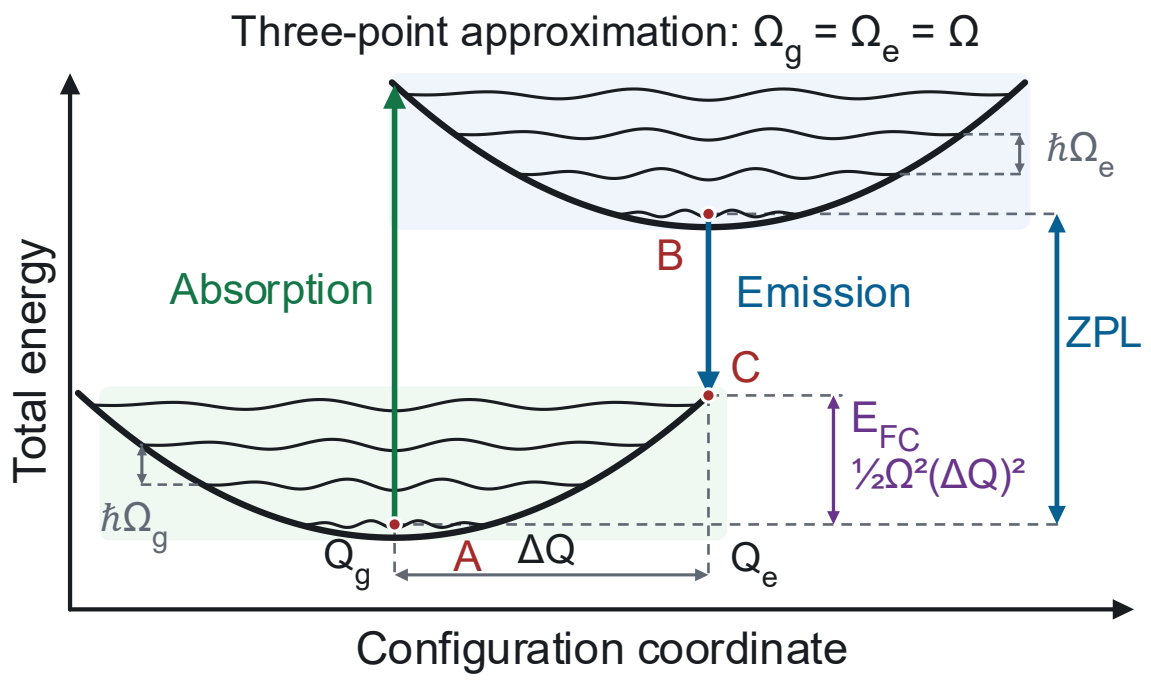

FIG. 1. Configuration-coordinate diagram for the three-point model. The ground- and excited-state equilibrium coordinates are $Q_g$ and $Q_e$, separated by ΔQ. A and B represent the relaxed ground and excited states; C represents the ground-state electronic configuration at $Q_e$. Their energies give $E_{ZPL} = E_B - E_A$ and $E_{FC} = E_C - E_A$. The effective frequencies are assumed equal, $\Omega_g = \Omega_e = \Omega$.

### C. Photoluminescence spectrum

We calculated the PL spectrum of $VV_a^{1+}$ with Py-Photonics using its default temperature and spectral broadening.[22] The ground- and excited-state geometries were obtained with HSE. The phonon force constants were calculated with the Perdew–Burke–Ernzerhof (PBE) functional using 1722 displaced structures of the 574-atom defect supercell.[23]

The structural displacement was projected onto the phonon eigenvectors to obtain the mode-resolved Huang–Rhys factors,

$$\Delta Q_k = \sum_{\alpha,i} \sqrt{m_\alpha}(r_{e,\alpha i} - r_{g,\alpha i}) e_{k,\alpha i}, \quad S_k = \frac{\omega_k |\Delta Q_k|^2}{2\hbar}. \tag{11}$$

where $e_{k,\alpha i}$ is the normalized eigenvector component of mode k and $\omega_k$ is its frequency. The electron–phonon coupling spectral density is

$$S(\hbar\omega) = \sum_k S_k \delta(\hbar\omega - \hbar\omega_k). \tag{12}$$

We obtained the PL lineshape with the generating-function formalism and determined the DWF from the ZPL contribution to the calculated emission spectrum.[24] Equation (4) gives an effective-mode estimate of this zero-phonon contribution within the three-point model.

## III. RESULTS AND DISCUSSION

### A. Electronic structure and stability of the optical excited states

Figs. 2–4 show the spin-resolved defect levels and the optical transitions considered for $V_C$, $VV_a$, and $VV_b$. The ground-state spins of $V_C^0$, $V_C^{1+}$, and $V_C^{1-}$ are $S_{spin}$ = 1, 1/2, and 3/2, respectively. For both divacancy configurations, the neutral, singly positive, singly negative, and doubly negative charge states have $S_{spin}$ = 1, 1/2, 3/2, and 2. Here $S_{spin}$ denotes the total electronic spin, to distinguish it from the Huang–Rhys factor S.

In its high-symmetry $C_{3v}$ configuration, $V_C^0$ has a nondegenerate $a_1$ level and a doubly degenerate e level within the band gap; the components of the latter are labeled $e_1$ and $e_2$ in Fig. 2(a). The singly occupied e manifold in the majority-spin channel is Jahn–Teller active, and the distortion splits the e-derived levels [Fig. 2(b)]. Excitation between these levels leads to an interchange of their ordering (degenerate exchange) after ΔSCF relaxation (see Fig. 2(c)). Likewise, the $a_1 \rightarrow e_1$ excitation of $V_C^{1+}$ [Fig. 2(d)] inverts the occupied and unoccupied levels (order interchange) after the ΔSCF relaxation as shown in Fig. 2(e). The imposed occupations therefore do not yield stable excited-state solutions for these transitions. Such changes in level ordering have also been reported in constrained-DFT studies of vacancies in SiC.[20, 21] For $V_C^{1-}$, the occupied and unoccupied levels maintain their ordering during excited-state relaxation, and the calculated ZPL energy is 3.013 eV (see Fig. 2(f)).

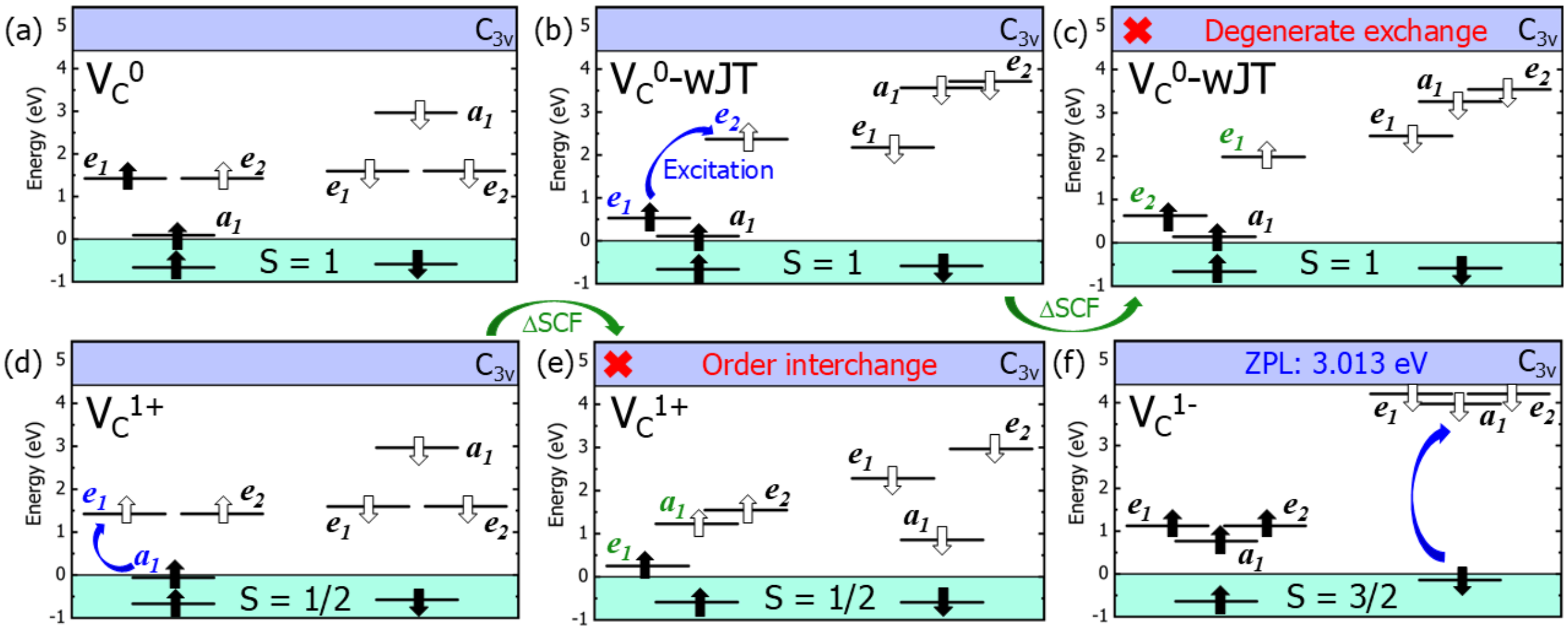

FIG. 2. Spin-resolved defect levels of the carbon vacancy. (a–c) $V_C^0$: the Jahn–Teller effect lifts the degeneracy of the e-derived levels in (b), labeled wJT (with Jahn–Teller), and excitation between these split levels leads to a level-order interchange (degenerate exchange) after ΔSCF relaxation in (c). (d,e) The $a_1 \rightarrow e_1$ excitation of $V_C^{1+}$ and the subsequent inversion of level ordering. (f) The optical transition of $V_C^{1-}$, with $E_{ZPL} = 3.013$ eV. Black and open arrows denote occupied and unoccupied states; blue curved arrows indicate the optical promotions. The shaded regions mark the host bands. The total electronic spin is labeled S in the panels.

The optical transitions of the axial divacancy depend strongly on its charge state. Those considered for $VV_a^0$ and $VV_a^{1-}$ undergo the level-order interchange as shown in Figs. 3(a) and 3(c), whereas $VV_a^{1+}$ and $VV_a^{2-}$ have stable constrained excited states. For $VV_a^{1+}$, which has $C_{2v}$ symmetry [Fig. 3(b)], the majority-spin $b_2 \rightarrow a_2$ transition gives a ZPL energy of 0.603 eV. The minority-spin transition of $VV_a^{2-}$ gives a ZPL energy of 2.074 eV [Fig. 3(d)]. Changing the charge state thus shifts the axial divacancy emission from the infrared to the visible range.

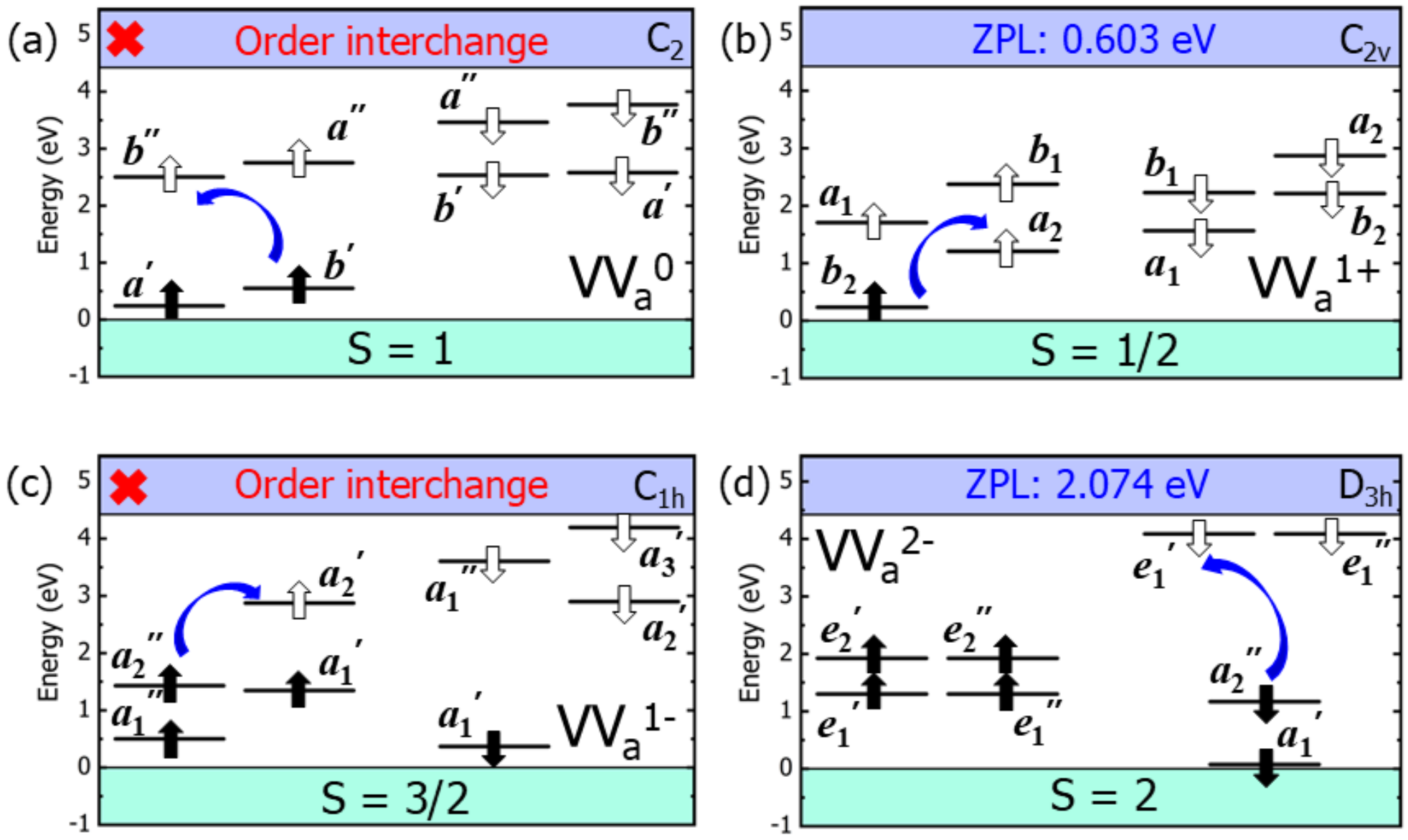

FIG. 3. Spin-resolved defect levels and optical transitions of the axial divacancy: (a) $VV_a^0$, (b) $VV_a^{1+}$, (c) $VV_a^{1-}$, and (d) $VV_a^{2-}$. The transitions in (a) and (c) exhibit a level-order interchange upon ΔSCF relaxation. The $b_2 \rightarrow a_2$ transition in (b) and the minority-spin transition in (d) have ZPL energies of 0.603 and 2.074 eV, respectively. The spin and symmetry labels are given in each panel; the graphical conventions follow Fig. 2.

For the basal divacancy, the transitions in $VV_b^{0}$ and $VV_b^{1+}$ undergo an order interchange of occupied and unoccupied levels as shown in Figs. 4(a) and 4(b), respectively. The singly negative and doubly negative charge states support the transitions in Figs. 4(c) and 4(d), with ZPL energies of 0.557 and 2.408 eV, respectively. The majority-spin $a_u \rightarrow b_g$ transition of $VV_b^{1-}$ provides a second infrared emitter for comparison with $VV_a^{1+}$. Thus, $V_C^{1-}$, $VV_a^{1+}$, $VV_a^{2-}$, $VV_b^{1-}$, and $VV_b^{2-}$ support stable excited-state configurations.

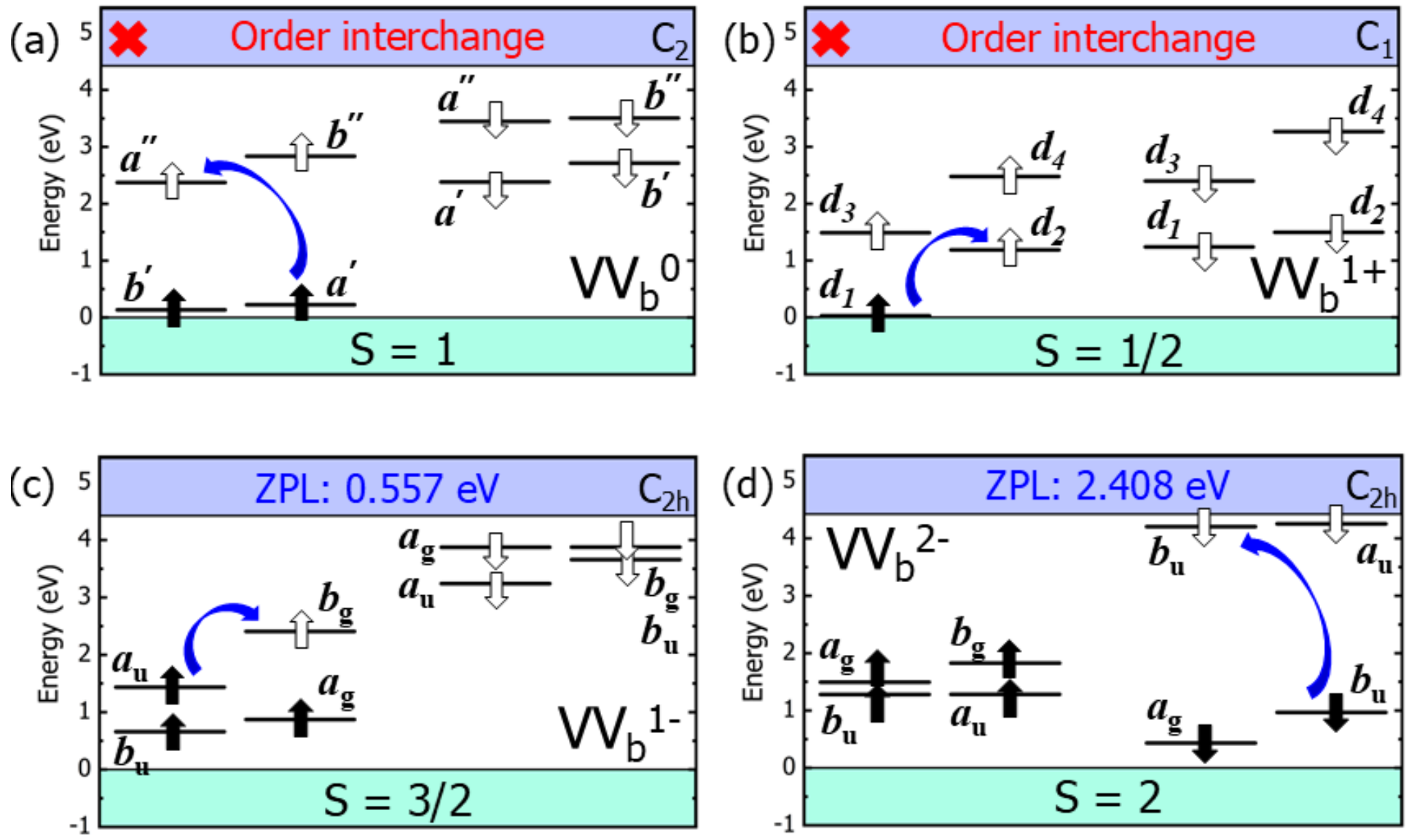

FIG. 4. Spin-resolved defect levels and optical transitions of the basal divacancy: (a) $VV_b^{0}$, (b) $VV_b^{1+}$, (c) $VV_b^{1-}$, and (d) $VV_b^{2-}$. The transitions in (a) and (b) exhibit a level-order interchange upon ΔSCF relaxation. The majority-spin $a_u \rightarrow b_g$ transition in (c) and the minority-spin transition in (d) have ZPL energies of 0.557 and 2.408 eV, respectively.

## B. Screening intrinsic defects for zero-phonon-line emission

We assessed the five intrinsic defects with stable excited-state configurations using the three-point configuration-coordinate model to calculate their optical parameters and radiative and nonradiative decay rates.[14] Table I lists these results alongside the vertical absorption and emission energies. The ZPL emission rate, $\Gamma_{ZPL}$, and ZPL quantum efficiency, $\eta_{ZPL}$, are used together to identify the most favorable emitter.

The doubly negative divacancies, $VV_a^{2-}$ and $VV_b^{2-}$, have weak optical transitions. Their ZPL energies of 2.074 and 2.408 eV lie in the visible range [Figs. 3(d) and 4(d)], but their transition dipole moments are $1.92 \times 10^{-4}$ and $1.24 \times 10^{-3}$ eÅ, respectively. With Huang–Rhys factors of 3.527 and 3.569, these small matrix elements give ZPL emission rates of $8.85 \times 10^{-8}$ and $5.53 \times 10^{-6}$ MHz as listed in Table I. Their radiative quantum efficiencies are close to unity because multiphonon decay is also slow. Their radiative lifetimes of $3.73 \times 10^{8}$ and $5.57 \times 10^{6}$ ns indicate very slow photon emission, making these defects less favorable as ZPL emitters.

$V_C^{1-}$ has the largest total radiative rate, $3.07 \times 10^{2}$ MHz, aided by its high transition energy of 3.013 eV (see Fig. 2(f)). Its structural displacement of $\Delta Q = 0.593$ amu$^{1/2}$ Å gives S = 3.300 and a DWF of 3.69%. Most radiative emission therefore occurs through the phonon sideband, leaving $\Gamma_{ZPL} = 1.24 \times 10^{1}$ MHz (see Table I). Although multiphonon decay is negligible in this model, phonon-sideband emission limits the ZPL quantum efficiency to 4.03%.

The infrared transitions of $VV_a^{1+}$ and $VV_b^{1-}$ involve much smaller structural displacements and have correspondingly larger DWFs. For $VV_a^{1+}$, the $b_2 \rightarrow a_2$ excitation in Fig. 3(b) gives $\Delta Q = 0.201$ amu$^{1/2}$ Å. Combined with an effective phonon energy of 81.50 meV, this displacement yields S = 0.395 and a DWF of 67.37%, the largest three-point DWF among the five centers. For $VV_b^{1-}$, $\Delta Q = 0.250$ amu$^{1/2}$ Å and S = 0.561 give a DWF of 57.06% as listed in Table I. Both infrared centers therefore emit a substantial fraction of their luminescence into the ZPL.

The vertical absorption and emission energies can also reflect lattice relaxation (see Table I). For $VV_a^{1+}$, $E_{abs} = 0.638$ eV and $E_{em} = 0.570$ eV lie close to the ZPL energy of 0.603 eV. Their separation of 0.068 eV is the smallest among the five centers, compared with 0.087 eV for $VV_b^{1-}$ and 0.427–0.552 eV for the other three defects. This small absorption–emission energy separation is consistent with the weak lattice relaxation and large zero-phonon contribution of $VV_a^{1+}$.

$VV_a^{1+}$ also has a larger transition dipole moment: $5.71 \times 10^{0}$ eÅ, more than three times the value of 1.68

$\times 10^0$ eÅ for $VV_b^{1-}$. Together with its larger DWF and slightly higher transition energy, this gives $\Gamma_{ZPL} = 4.39 \times 10^1$ MHz, compared with $2.53 \times 10^0$ MHz for $VV_b^{1-}$ (see Table I). The ZPL emission rate of $VV_a^{1+}$ is therefore approximately 17 times that of $VV_b^{1-}$.

TABLE I. Calculated properties of intrinsic color centers in hexagonal diamond. $E_{abs}$ and $E_{em}$ denote the vertical absorption and emission energies, respectively. The three-point quantities use n = 2.4 and $w_{e,g} = 0.1$ eV amu$^{-1/2}$ Å$^{-1}$. $E_{FC}$ is obtained from $S\hbar\Omega$, and $\lambda_{ZPL} = hc/E_{ZPL}$. The QFC row indicates whether quantum frequency conversion is required to reach the telecom range.

| **Parameter** | **Unit** | $V_C^{1-}$ | $VV_a^{1+}$ | $VV_a^{2-}$ | $VV_b^{1-}$ | $VV_b^{2-}$ |
|---|---|---|---|---|---|---|
| Total spin $S_{spin}$ | | 3/2 | 1/2 | 2 | 3/2 | 2 |
| Point group | | $C_{3v}$ | $C_{2v}$ | $D_{3h}$ | $C_{2h}$ | $C_{2h}$ |
| g | | 1 | 1 | 3 | 1 | 1 |
| $E_{ZPL}$ | eV | 3.013 | 0.603 | 2.074 | 0.557 | 2.408 |
| $E_{abs}$ | eV | 3.306 | 0.638 | 2.331 | 0.602 | 2.630 |
| $E_{em}$ | eV | 2.754 | 0.570 | 1.847 | 0.515 | 2.203 |
| $\lambda_{ZPL}$ | nm | 411.5 | 2057.3 | 597.8 | 2227.2 | 515.0 |
| $\Delta Q$ | amu$^{1/2}$ Å | 0.593 | 0.201 | 0.676 | 0.250 | 0.722 |
| $\hbar\Omega$ | meV | 78.44 | 81.50 | 64.43 | 74.83 | 57.21 |
| $E_{FC}$ | meV | 258.87 | 32.19 | 227.21 | 41.98 | 204.19 |
| $\lvert\mu\rvert$ | eÅ | $1.16 \times 10^0$ | $5.71 \times 10^0$ | $1.92 \times 10^{-4}$ | $1.68 \times 10^0$ | $1.24 \times 10^{-3}$ |
| S | | 3.300 | 0.395 | 3.527 | 0.561 | 3.569 |
| DWF | % | 3.69 | 67.37 | 2.94 | 57.06 | 2.82 |
| $\Gamma_R^0$ | MHz | $3.36 \times 10^2$ | $6.51 \times 10^1$ | $3.01 \times 10^{-6}$ | $4.43 \times 10^0$ | $1.96 \times 10^{-4}$ |
| $\Gamma_R$ | MHz | $3.07 \times 10^2$ | $6.16 \times 10^1$ | $2.68 \times 10^{-6}$ | $4.10 \times 10^0$ | $1.80 \times 10^{-4}$ |
| $\Gamma_{PSB}$ | MHz | $2.95 \times 10^2$ | $1.78 \times 10^1$ | $2.59 \times 10^{-6}$ | $1.57 \times 10^0$ | $1.74 \times 10^{-4}$ |
| $\Gamma_{ZPL}$ | MHz | $1.24 \times 10^1$ | $4.39 \times 10^1$ | $8.85 \times 10^{-8}$ | $2.53 \times 10^0$ | $5.53 \times 10^{-6}$ |
| $\Gamma_{NR}$ | MHz | $2.48 \times 10^{-17}$ | $3.11 \times 10^2$ | $1.45 \times 10^{-9}$ | $2.83 \times 10^3$ | $1.06 \times 10^{-19}$ |
| $\tau_{rad}$ | ns | $3.25 \times 10^0$ | $1.62 \times 10^1$ | $3.73 \times 10^8$ | $2.44 \times 10^2$ | $5.57 \times 10^6$ |
| $\tau_{total}$ | ns | $3.25 \times 10^0$ | $2.68 \times 10^0$ | $3.73 \times 10^8$ | $3.53 \times 10^{-1}$ | $5.57 \times 10^6$ |
| $\eta_{rad}$ | % | 100 | 16.5 | 99.9 | 0.144 | 100 |
| $\eta_{ZPL}$ | % | 4.03 | 11.8 | 3.30 | 0.0892 | 3.08 |
| QFC | | Yes | Yes | Yes | Yes | Yes |

Nonradiative decay further distinguishes the two infrared centers. Their low ZPL energies favor multiphonon relaxation, consistent with the energy dependence discussed by Turiansky *et al.* and Iwamoto *et al.* [6, 14] For $VV_a^{1+}$, $\Gamma_{NR} = 3.11 \times 10^2$ MHz and $\Gamma_R = 6.16 \times 10^1$ MHz give $\eta_{rad} = 16.5\%$ and $\eta_{ZPL} = 11.8\%$. Its radiative and total excited-state lifetimes are $1.62 \times 10^1$ and $2.68 \times 10^0$ ns, respectively. For $VV_b^{1-}$, the higher nonradiative rate of $2.83 \times 10^3$ MHz and lower radiative rate of $4.10 \times 10^0$ MHz reduce $\eta_{ZPL}$ to 0.0892%, despite its large DWF.

$VV_a^{1+}$ is the most promising ZPL emitter among the five candidates, with both the largest $\Gamma_{ZPL}$ and the highest $\eta_{ZPL}$. Its small structural displacement suppresses phonon-sideband emission, and its large transition dipole supports a strong infrared ZPL.

### C. Photoluminescence lineshape of $VV_a^{1+}$ from a multimode phonon calculation

The three-point configuration-coordinate model describes lattice relaxation during the optical transition with a single effective phonon mode. To resolve the phonon sideband and the contributions of individual modes, we calculate the multimode PL lineshape of $VV_a^{1+}$. The spectral density in Fig. 5(a) is largest near 75–80 meV, close to the effective phonon energy of 81.50 meV in Table I, with additional features near 50, 115, and 145 meV. Several modes therefore contribute to optical relaxation, but the dominant coupling lies in the energy range captured by the three-point model. Summing the mode-resolved contributions gives a total Huang–Rhys factor of 0.366 [Fig. 5(a)], close to the three-point value of 0.395 as listed in Table I.

The calculated PL spectrum in Fig. 5(b) is dominated by the ZPL at approximately 0.603 eV, with a much weaker phonon sideband at lower energies. The broad feature near 0.52–0.53 eV is consistent with emission accompanied by a phonon in the dominant 75–80 meV range of Fig. 5(a). The multimode DWF of 69.37% [Fig. 5(b)] is close to the three-point value of 67.37% (see Table I). The agreement in both the Huang–Rhys factor and DWF for this center supports the three-point model for comparative screening.

The DWF of 69.37% indicates much weaker electron–phonon coupling in $VV_a^{1+}$ than in $NV^{1-}$ in cubic diamond, whose measured low-temperature DWF is approximately 3.2%.[24] The calculated DWF for $VV_a^{1+}$ is comparable to the greater than 70% ZPL emission reported for $SiV^{1-}$, whose ZPL lies near 737 nm.[5] $VV_a^{1+}$ thus combines a large zero-phonon contribution with a substantially longer emission wavelength of approximately 2057.3 nm. As an intrinsic divacancy, it offers these optical properties without impurity incorporation into the host.

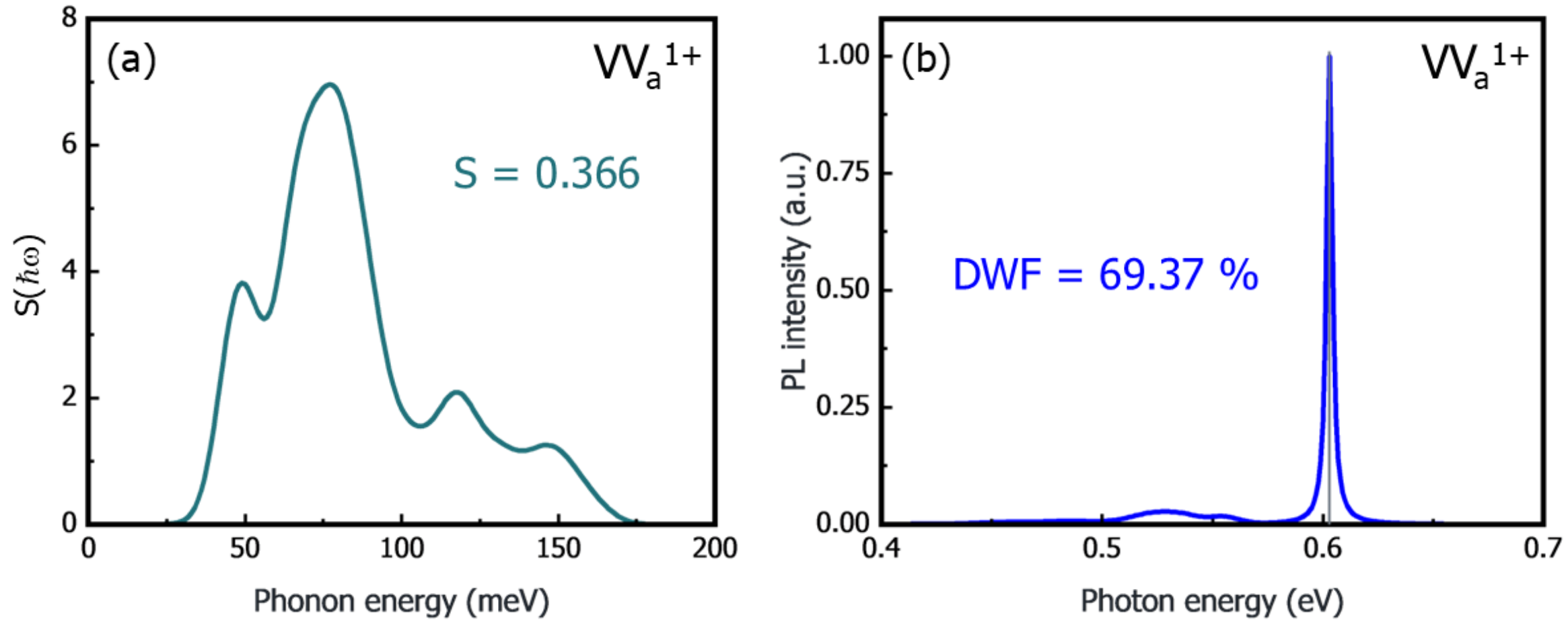

FIG. 5. Electron–phonon coupling and photoluminescence of $VV_a^{1+}$. (a) Huang–Rhys spectral density S(ħω), with a total Huang–Rhys factor of 0.366 and a dominant contribution near 75–80 meV. (b) Normalized PL spectrum as a function of emitted photon energy. The ZPL lies near 0.603 eV, and the phonon sideband extends to lower energies. The calculated DWF is 69.37%.

### D. Emission wavelength and quantum frequency conversion

Emission wavelength is another criterion for selecting $VV_a^{1+}$. We use 1310 nm in the O band and 1550 nm in the C band as representative targets within the 1260–1675 nm telecom range considered by Turiansky et al.[6] The infrared ZPLs occur at 2057.3 nm for $VV_a^{1+}$ and 2227.2 nm for $VV_b^{1-}$, while the other three transitions lie between 411.5 and 597.8 nm (Table I). $VV_a^{1+}$ has the smallest photon-energy separation from both targets among the five candidates: 0.344 eV at 1310 nm and 0.197 eV at 1550 nm. Together with its ZPL emission rate and quantum efficiency and its multimode PL spectrum, this spectral proximity makes $VV_a^{1+}$ a promising infrared color center for fiber-based quantum communication.

All five ZPLs lie outside the telecom range, so each candidate would require QFC to reach the O or C band (see Table I).[6] Diamond $NV^{1-}$ photons have been converted from 637 to 1588 nm while preserving their single-photon character, and telecom conversion has enabled SiV-based entanglement over deployed fiber.[4, 25] Reaching 1310 or 1550 nm would require upconversion for the two infrared transitions and downconversion for the other three. The strong ZPL emission and large DWF of $VV_a^{1+}$ favor its use in a telecom interface; the conversion efficiency and added noise would depend on the nonlinear optical implementation.

## IV. CONCLUSIONS

Our first-principles screening identifies $VV_a^{1+}$ as the most promising ZPL emitter among five intrinsic defects in bulk hexagonal diamond with stable optical excited states. It has the highest calculated ZPL emission rate (43.9 MHz) and ZPL quantum efficiency (11.8%), together with a multimode Debye–Waller factor of 69.37%. Its native 2057.3 nm emission is also closest in photon energy to the representative telecom targets. Upconversion to 1310 nm in the O band or 1550 nm in the C band would require photon-energy increases of 0.344 and 0.197 eV, respectively. These optical and spectral properties make $VV_a^{1+}$ a promising intrinsic emitter for fiber-based quantum communication through suitable quantum frequency conversion.

## ACKNOWLEDGMENTS

X. Zhang and H. Pei contribute equally to this work. This work was supported by the National Natural Science Foundation of China (NSFC) under grant Nos. 12304110 and 12404093.

## DATA AVAILABILITY

All data generated or analyzed during this study are included in this published article.